\documentclass[12pt,amsonts,hyperref]{iopart}
\usepackage{iopams}
\usepackage{epsfig}
\usepackage{hyperref}
\def\({\left(}
\def\){\right)}
\def\[{\left[}
\def\]{\right]}
\def\vect#1{\skew{-1}{\mathaccent"017E}{#1}}
\def\vp{\vect p}
\def\vx{\vect x}

\def\tmp#1{}
\def\be{\begin{equation}}
\def\e{\end{equation}}
\def\ee{\end{equation}}
\begin{document}

\title{Casimir type effects for scalar fields interacting with material slabs}

\author{I V  Fialkovsky\dag\ddag, V N  Markov \P \ and  Yu M
Pis'mak\dag}
\address
{\dag \ Department of Theoretical Physics,  State University of
Saint-Petersburg, Russia}
\address
{\ddag \ Insituto de F\'isica, Universidade de S\~ao Paulo, S\~ao Paulo,  Brazil}
\address
{\P \ Department of Theoretical Physics, Petersburg Nuclear
Physics Institute, Russia} \ead{\dag ifialk@gmail.com,
pismak@jp7821.spb.edu, \P markov@thd.pnpi.spb.ru}


\begin{abstract}
We study the field theoretical model of a  scalar field in
presence of spacial inhomogeneities in the form of one and two finite width mirrors
(material slabs). The interaction of the scalar field with the
defect is described with position-dependent mass term.

For a single layer system we develop a rigorous calculation method
and derive explicitly the propagator of the theory, the S-matrix
elements and the Casimir self-energy of the slab.
Detailed investigation of particular limits of self-energy
is presented, and connection to known cases is discussed.

The calculation method is found applicable to the two mirrors case as well. With its help we
derive the corresponding Casimir energy and analyze it. For particular values of
parameters of the model an obtained result recovers the Lifshitz formula. We also propose a
procedure to unambiguously obtain  the finite Casimir \textit{self}-energy of a single slab without reference to any renormalization conditions. We hope that our approach can be applied to calculation of Casimir self-energies in other demanded cases (such as dielectric ball, etc.)

\end{abstract}


\section{Introduction}
Usual models of interaction of elementary particles are constructed in a framework of the
quantum field theory (QFT) in the homogenous infinite space-time \cite{BogShir}.
However, as such they can not be applied for description of phenomena of interaction
of quantum fields with macroscopic bodies.
At least the form of latter  must be presented in the model,
and it can change essentially both the spectrum and dynamics of
the excited states of the model as well as the properties of the
ground (vacuum) state of the system.

First quantitative results for such effects were obtained by H. Casimir in 1948. He predicted
\cite{Casimir'48} macroscopical attractive force between two
uncharged conducting plates placed in vacuum. The force appears
due to the influence of the boundary conditions on the
electromagnetic quantum vacuum fluctuations. Nowadays the Casimir
effect is verified by experiments with a precision of $0.5$\%
(see \cite{Klimchitskaya 05} for a review).

The properties of vacuum fluctuations in curved spaces,
the scalar field models with various boundary
conditions and their application to the description of real
electromagnetic effects were actively studied throughout the last
decades (see discussion and references in \cite{Blau Visser Wipf 88}-\cite{Milton-BOOK'01}).
However, it was well understood from the beginning that boundary conditions must be
considered just as an approximate description of a complex
interaction of quantum fields with the matter. A generalization of
the boundary conditions method has been proposed by Symanzik
\cite{Symanzik'81}. In the framework of path integral formalism he
showed that the presence of material boundaries (two dimensional
defects) in the system can be modeled with a surface term added to
the action functional. Such singular potentials with
$\delta$-type profile functions concentrated on the defect surface
reproduce some simple boundary conditions (namely Dirichlet and
Neumann ones) in the strong coupling limit. The additional action
of the defect should not violate basic principles  of the bulk
model such as gauge invariance (if applicable), locality and
renormalizability.

The QFT systems with $\delta$-type potentials are mostly investigated
for scalar fields, see for instance \cite{Milton deltas 04}.
In \cite{Markov Pismak'05}--\cite{FMP 08} the
Symanzik's approach was used to describe similar
problems in complete quantum electrodynamics (QED), and all
$\delta$-potentials consistent with the QED basic principles were
constructed.

It seems quite natural to try applying the same method for
the description of the interaction of quantum fields with bulk macroscopic
inhomogeneities (slabs, finite width mirrors, etc.) and to study
Casimir effects in systems of such kind.
The Symanzik's method of adding extra potential terms into the action of a system
was used to model
the interaction of quantum scalar fields with bulk defects in a number of
papers
(e.g. \cite{Bordag'95}, \cite{Graham Jaffe 04}-%
\cite{Cavero-Pelaez Milton Wagner 05}, and others). However, most
of them were devoted to study of a limiting procedure of
transition from a bulk potential of the defect to the surface
$\delta$-potential one, without paying much attention to other properties.

Traditionally, there are several approaches to study electromagnetic Casimir
effect between material slabs. The electromagnetic field and the material
bodies can be treated macroscopically  and by employing a dissipation fluctuation
theorem one  obtains the field correlation functions that are needed to construct
the Maxwell stress tensor. In alternative approach the presence of dielectric bodies
is described by means of a spatially varying permittivity that is a complex function
of the frequency. Within these approaches the influence of dielectric and geometrical
properties on the Casimir force between dispersing and absorbing plates has been studied
in \cite{Kup,Raabe}. A dependence of the Casimir force on thickness of the slabs was
considered in \cite{Pirozhenko}. A calculation of the  Casimir force in a (multi)layer
system have been also performed in \cite{modes,fields,Ellingsen,Sirvent}. Most recently
there was an attempt to construct  quantum electrodynamics in the presence of dielectric
media (i.e. volume defects of special kind) \cite{Eberlein Robaschik 05}. From the field
theoretical point of view none of these methods were truly successful. Particularly  the
celebrated Lifshitz approach \cite{Lifshitz'56} to the description of the interaction between
dielectric bodies still attracts a lot of arguments if dispersion is present in the system
and formally is proved only for the interaction over the vacuum gap, see \cite{Pitaevskii 09}.
While in some approaches there could be no action functional constructed
\cite{Bordag-Mohideen-Mostepanenko-OBZOR'01}, there are also  general severe
problems in construction of path-integral quantization for the QED
systems with dielectrics \cite{Bordag'98} and calculation of the heat kernel
coefficients for such systems \cite{Vassilevich 03}.
Moreover, as yet remains unresolved the mystery of Casimir energy of a dielectric ball which is
defined unambiguously only in the dilute limit or for the speed of light being constant across
the boundary \cite{Bordag'99}.

On the other hand, existing results for the Casimir energy of a scalar
field in presence of a single planar layer of finite width $\ell$
are contradictory. The formulae presented recently in \cite{Fosco
Lombardo Mazzitelli 08} do not coincide with previous
calculations made in \cite{Bordag'95} as discussed in \cite{FMP Comm 08}. The only attempt to
calculate the propagator in such system was undertaken in
\cite{Aguiar 93}, where hardly any explicit formulae were after all
presented. The system of two interacting slabs has never been considered
within QFT models.

Thereby, one can see that the specificity of finite volume effects generated
by inhomogeneities in QFT has not been yet adequately explored. Our work is
dedicated to clarify the problem, and to solve existing controversy within an
accurate and unambiguous approach. On the other hand, considered type of square well
potentials has its own range of applicability including transport in graphene
and other condensed matter systems \cite{cond-mat}, and (apparently) superluminal
effects in wave propagation \cite{super-lum}. While consideration of the scalar fields
may only appear restricting, for the case of planar geometries it is apparently sufficient.
It is due to the well known fact that in this case the theory of
electromagnetic (EM) field is equivalent to two scalar field theories as $TE$ and $TM$ modes of the former field do not mix.
The study of EM fields in the geometries, when such decomposition does not appear,
will be the scope of the future work.

Thus, we consider a model of massive scalar field interacting with the volume defects
--- finite width material slabs or mirrors, --- whose properties are effectively
described by single macroscopical parameter, the coupling constant. By construction the model is renormalizable in the sense applicable to QFT with spatial inhomogeneities given in \cite{Symanzik'81}.
Working with renormalizable models has an important advantage that their long distance (macroscopical)
properties such as Casimir interaction can be described without explicit knowledge of microscopical
structure even if there is a true physical cut-off present in the system, \cite{Zinn-Justin}. Such description can be given in terms of just a few effective parameters. In our case, the role of these parameters is played by the coupling constant. Any physical cut-off dependence is disregarded in light of renormalizability of the model.


We develop a  purely field theoretical approach and construct
explicitly the propagator for the system with a single finite width mirror (Sect.
\ref{1L pr}) and the S-matrix elements (Sect. \ref{1L Sc}). Based on these new results
in Section \ref{E single Sect}, we recover the known expression for the self-interaction
of a single slab and supply it with the investigation of its asymptotical behavior in different
physically interesting regimes (Sections \ref{Dir lim Sect}--\ref{no div}).

In Section \ref{2L} the Casimir energy of two
interacting finite width slabs
described within the scalar field model with step potential
is obtained for the first time (Sect. \ref{2L en sect})
and several limiting cases are considered (Sect. \ref{Limits sect}).
We also reveal its correspondence to other known results in the literature, and
in section \ref{Lifsh sect} discuss its connection to the Lifshitz formula. Possible ways to
construct divergence free limits for Casimir self-interaction of
a solitary body are presented in Sect. \ref{Self sect}. In the Conclusion the
main results are summarized. In the \ref{Reg depen} we briefly discuss a by-side
question of comparison of different regularizations within QFT.

\section{Statement of the problem}
Let us consider a model of a scalar field interacting with a
space defect with nonzero volume. Using the Symanzik's approach, we describe
such system by the action functional with an additional mass term being non-zero only
inside the defect
\begin{eqnarray}
  S &=& S_0+S_{\it def}  \label{action}\\
    && S_0=\frac12\int d^4x \phi(x) (-\partial^2_x+m^2) \phi(x)
        \nonumber\\
    &&
       S_{\it def}=\frac\lambda2\int d^4x  \theta(\ell, x_3)\phi^2(x)
       \nonumber
\end{eqnarray}
where $\partial_x^2=\partial^2/\partial
x_0^2+\ldots+\partial^2/\partial x_3^2$%
\footnote{We operate in Euclidian version of the theory which
appears to be more convenient for the calculations.}).
In the simplest case, the defect could be considered as a homogenous
and isotropic infinite plane layer of the thickness $\ell$ placed
in the $x_1x_2$ plane. In this case it is sometimes called `piecewise
constant potential'. The distribution function $\theta(\ell,
x_3)$ is then equal to $1/\ell$ when $|x_3|<\ell/2$, and is zero
otherwise. In terms of the Heaviside step-function it can be
written as
\be \theta(\ell, x_3)\equiv
    [\theta(x_3+\ell/2)-\theta(x_3-\ell/2)]/\ell.
    \label{theta}
\ee
In the framework of QFT such potentials were approached for the
first time in \cite{Bordag'95}, and later in \cite{Feinberg Mann
Revzen 99}-\cite{Fosco Lombardo Mazzitelli 08}. However, the
latest consideration \cite{Fosco Lombardo Mazzitelli 08} was
proved to be inconsistent \cite{FMP Comm 08}.

To describe all physical properties of the systems it is
sufficient to calculate the generating functional for the Green's
functions
\be
    G[J]= N\int D\phi\, \exp\{-S[\phi]+J\phi\},
        \quad N^{-1}=\int D\phi\, \exp\{-S_0[\phi]\}
        \label{G(J)}
\e
where $J$ is an external source, and  normalization for the
generating functional has been chosen in such a way that
$G[0]|_{\lambda=0}=1$.

Introducing in (\ref{G(J)}) an auxiliary field $\psi$ defined in the
volume of the defect only, we can present the defect  contribution
to $G[J]$ as
\be
 \exp\left\{-\frac{\lambda}{2\ell}\int d\vx\int_{-\ell/2}^{\ell/2} dx_3 \phi^2(x) \right\}=
    \label{psi}
   \ee
   $$
      =C \int D\psi \exp\left\{\int d\vx\int_{-\ell/2}^{\ell/2} dx_3\(-\frac{\psi^2}{2}
                                    +i\sqrt{\kappa}\psi\phi\)\right\}
 $$
which is written for the case of a single layer.
$C$ is an appropriate normalization constant,
and $\kappa \equiv {\lambda}/{\ell}$, here and below we
denote with arrows the $(0,1,2)$ components of 4-vectors, i.e.
$\vx=(x_0, x_1,x_2)$. When generalized
to other defects the $x_3$-integration in
(\ref{psi}) should be performed only within their support.

With help of projector onto the volume of the layer ${\cal
O}=\theta(x_3+\ell/2)-\theta(x_3-\ell/2)$ acting as
$
    \psi{\cal O}\phi\equiv \int d\vx\int_{-\ell/2}^{\ell/2} dx_3\psi\phi
$,
we can perform the functional integration over $\phi$, and
consequently over $\psi$. As the result we get
\be G[J]= [{\rm Det}Q]^{-1/2} e^{\frac12J\hat S_L J}, \quad
    \hat S_L=D-\kappa (D{\cal O}) Q^{-1}({\cal O} D),
    \label{G_fin}
\e \be
    Q=\textbf{1}+\kappa({\cal O} D{\cal O}).
    \label{Qx}
\e
Here the unity operator $\textbf{1}$, as well as the whole $Q$, is
defined in the volume of the defect only
$(-\ell/2,\ell/2)\times{\mathbb R}^3$, and
$D=(-\partial^2+m^2)^{-1}$  is  the standard (Feynman) propagator
of the free scalar field. We shall note here that the outlook of
(\ref{G_fin}) completely coincides with the expression for the generating
functional $G[J]$ in the case of delta-potential defect instead of
patchwise constant one subject to appropriate redefinition of the
projecting operator $\mathcal{O}$. It is also evident that a
straightforward generalization is possible for non-constant
$\kappa$($=\lambda/\ell$) with $\lambda$ depending on $x_3$.

\section{One mirror system}
\label{single}

\subsection{Calculation of the propagator}\label{1L pr}
To calculate the propagator $\hat{S}$  defined according to
(\ref{G_fin}) let us first derive an explicit formula for the
operator $W\equiv Q^{-1}$.

For this purpose we make the Fourier transformation on
the  coordinates parallel to the defect (i.e.  $x_0$, $x_1$,
$x_2$). In such mixed $\vp$-$x_3$ representation the propagator
$D$ of the system without a defect is given by
\be D(x)=\int\frac{d^3\vp}{(2\pi)^3}e^{i \vp \vx}\, {\cal
D}_{E^2}(x_3),
    \label{fourier}
    \qquad {\cal D}_{V}(x)\equiv\frac{e^{-\sqrt{V}|x|}}{2\sqrt{V}}
\ee
with $E=\sqrt{p^2+m^2}$, $\vp=(p_0,p_1,p_2)$. Then $W$ can be defined through the
following operator equation
\be
    W + \kappa {\cal D}_{E^2} W =1.
    \label{W}
\ee
By construction the free scalar propagator  ${\cal
D}_{E^2}(x,y)\equiv{\cal D}_{E^2}(x-y)$ is the Green's function of
the following ordinary differential operator
\be
    K_{V}(x,y)=\left(-\frac{\partial^2}{\partial x^2}+V\right)\delta(x-y)
    \label{s0}
\ee
for $V=E^2$. Multiplying both sides of (\ref{W}) with $K_{E^2}$
and using obvious relation $K_V-K_{V'}=V-V'$ we get
\begin{equation}
    K_\rho U = -\kappa
    \label{s1}
\end{equation}
where  $\rho\equiv \kappa+E^2$ and $U\equiv W-1$.

The general solution to this (inhomogeneous) operator equation can
be written as a sum of its partial solution and the general
solution of its homogeneous version. Taking into account the
symmetry property of $U$: $U(x,y)=U(y,x)$, which follows from its
definition, we arrive at
\be
U(x,y)=-\kappa {\cal D}_{\rho}(x,y)
    +a e^{(x+y) \sqrt{\rho}}+b\left(e^{(x-y) \sqrt{\rho}}+e^{(y-x) \sqrt{\rho}}\right)
        +c e^{-(x+y) \sqrt{\rho}}
    \label{U}
\ee
where $a$, $b$ and $c$ are some constants. To derive them we
introduce $W=1+U$ into (\ref{W}) and require its identical
validity for all $x$ and $y$. It yields
\be a=c=-\frac{\xi \kappa^2 e^{\ell
    \sqrt{\kappa+E^2}}} {2 \sqrt{\kappa+E^2}},\qquad
b=-\frac{\xi
    \kappa (E-\sqrt{\kappa+E^2})^2}
    {2 \sqrt{\kappa+E^2}},
           \label{abc}
\ee
$$
\xi=\frac1{e^{2\ell\sqrt{\kappa+E^2}}(E+\sqrt{\kappa+E^2})^2-(E-\sqrt{\kappa+E^2})^2}.
$$

Substituting $Q^{-1}=1+U$ into (\ref{G_fin}) and using (\ref{abc})
we can finally derive the explicit formulae for the modified
propagator of the system $\hat S\equiv\hat S (\vp,x_2,y_3)$
\be
\hat S(\vp, x_3,y_3)=\left\{%
\begin{array}{ll}
    S_{--},         & x_3<-\ell/2,\ y_3<-\ell/2\\
    S_{-\circ},     & x_3<-\ell/2,\ y_3\in(-\ell/2,\ell/2) \\
    S_{-+},         & x_3<-\ell/2,\ y_3>\ell/2\\
    S_{\circ \circ}, & x_3\in(-\ell/2,\ell/2),\ y_3\in(-\ell/2,\ell/2) \\
\end{array}%
\right.
\label{hat S}
\ee
where
\hskip-2cm\begin{eqnarray}
  \hskip-2.5cm
  &S_{--}& = \frac{e^{-E|x_3-y_3|}}{2E}
    +\frac{\kappa \xi }
        {2 E }e^{E (\ell+x_3+y_3)}(1-e^{2 \ell\sqrt{\kappa+E^2}})
    \label{S --}\\
  \hskip-2.5cm
  &S_{-\circ}&
   =
        \xi \({ e^{ (\ell/2+y_3)\sqrt{\kappa+E^2}}(\sqrt{\kappa+E^2}-E)+e^{(3\ell/2-y_3)\sqrt{\kappa+E^2} }(\sqrt{\kappa+E^2}+E)}\)
            e^{E(x_3+\ell/2)}\\
  \hskip-2.5cm
  &S_{-+} &= {2 \xi \sqrt{\kappa+E^2} e^{(\sqrt{\kappa+E^2}+E)\ell+E(x_3-y_3)}}
        \\
  \hskip-2.5cm
  &S_{\circ \circ} &= \frac{\xi e^{\ell\sqrt{\kappa+E^2}}}{2\sqrt{\kappa+E^2} }\(
        2\kappa \cosh[(x_3+y_3)\sqrt{\kappa+E^2}]+ \right.
        \label{S circ circ}\\
  \hskip-2.5cm
  &&    + \left.(\sqrt{\kappa+E^2}-E)^2 e^{\sqrt{\kappa+E^2} (|x_3-y_3|-\ell)}
            +(E+\sqrt{\kappa+E^2})^2 e^{\sqrt{\kappa+E^2} (\ell-|x_3-y_3|)} \)\nonumber.
\end{eqnarray}
We have divided the general expression of the propagator into four
parts according to the position of $x_3$, $y_3$ relative to the
defect, and all other cases could be easily derived using the
symmetry properties of the propagator.

Despite that the problem of square-well potential has been studied
extensively in quantum mechanics,  its
field theoretical interpretation was lacking. To the best of our
knowledge the only attempt to calculate the full propagator for
such system was presented in \cite{Aguiar 93}. However, the final
explicit expression is lacking there, but instructions for its construction are
given to the reader. Following these instructions one can compare
\cite{Aguiar 93} with (\ref{S --})-(\ref{S circ circ}) and find
them coinciding up to the Wick rotation. One should also note that in
Euclidian version of the problem the difference between under- and
above-barrier scattering vanishes. The $S_{\circ \circ,--}$ for 2-dimensional
problem were also presented in \cite{Milton-OBZOR'04}.

Fixing position of one of the interfaces of the potential and taking the
infinite width limit transforms the potential into a
step-function one.
The propagator for non-relativistic problem
with such potential was calculated
earlier, e.g. \cite{Grosche 93}. Subject to necessary redefinitions of
parameters, our results (\ref{S --})-(\ref{S circ circ}) for $\ell\to\infty$ coincide
with ones presented there.

\subsection{Scattering on the slab}\label{1L Sc}
In the quantum field theory the scattering processes are described by
the S-matrix elements between
different states.
To construct them for one-particle states
of particular impulses, we consider the full propagator
(\ref{hat S}) in real (Minkowski) space, and cut the external lines of the
free field propagator \be
    H(x,y)={D}^{-1}(x,z') \hat S(z',z'') {D}^{-1}(z'',y).
\ee
The defect potential depends only on $x_3$ thus both the energy
($p_0$) and impulse parallel to the defect ($\vp$) are conserved.
We omit corresponding delta-functions in the Fourier representation.
In the transverse direction we have
\begin{eqnarray*}
    H(p_3,q_3)&=&\int dx_3 dy_3\, e^{ip_3x_3+iq_3y_3} H(x_3,y_3)\\
        &=&-\kappa\int_{-\ell/2}^{\ell/2} dx_3 \int_{-\ell/2}^{\ell/2} dy_3\,
            e^{ip_3x_3+iq_3y_3}\(\delta(x_3-y_3)+U^M(x_3,y_3)\)
\end{eqnarray*}
where the last equation is due to the definition of $\hat S$ and
the obvious property that ${\cal D}_{E^2}^{-1}(x,z) {\cal
D}_{E^2}(z,y) {\cal O}(y) =\delta(x-y){\cal O}(y)$ is only
non-zero when both its arguments lie in the defect region
($x,y\in(-\ell/2,\ell/2)$). By $U^M$ we denote the $U$ operator (\ref{U}), (\ref{abc})
subject to the inverse Wick rotation.

Further we note, that in the Minkowski space the on-shell condition
leaves only two possible values for transversal impulse $p_3$:
$
    p_3=\pm \verb"E"$, $\verb"E"\equiv\sqrt{p_0^2-\vp^2-m^2}
$.
Due to the symmetry reasons there are only two independent values
of $H$ depending on the sign of $p_3,q_3=\pm \verb"E"$ . Thus the
non-trivial $S$-matrix elements depend only on one parameter
$p\equiv\verb"E"$ and are given by
\begin{eqnarray}
    h_f=-2ip-\frac{8p^2 \sqrt{\kappa-p^2} e^{\ell(ip+\sqrt{\kappa-p^2})}}
        {(e^{2\ell \sqrt{\kappa-p^2}}-1)(\kappa-2p^2)+2ip\sqrt{\kappa-p^2}(1+e^{2\ell \sqrt{\kappa-p^2}})}\\
    h_b=  \frac{2 i p \kappa (1-e^{2\ell \sqrt{\kappa-p^2} } )e^{i p \ell}}
        {(e^{2\ell \sqrt{\kappa-p^2}}-1)(\kappa-2p^2)+2ip\sqrt{\kappa-p^2}(1+e^{2\ell \sqrt{\kappa-p^2}})}.
\end{eqnarray}
where $h_f\equiv H(p,-p)$ is the forward scattering amplitude, and
$h_b\equiv H(p,p)$ --- the reflection one.

It is possible to check that the presented $S$-matrix amplitudes
satisfy the following analog of the optical theorem
\be
    2{\textrm{ Im}}(h_f)=-\frac{1}{2p}(|h_f|^2+|h_b|^2),\qquad
    2{\textrm{ Im}}(h_b)=-\frac{1}{2p}(h_f h_b^* + h_f^* h_b)
\e

\subsection{Casimir Energy}\label{E single Sect}
The Casimir energy density per unit area of the defect $S$ can be
presented with the following relation
\be {\cal E}=-\frac{1}{TS}\ln G[0]
    =\frac1{2TS}\,\mbox{Tr}\ln[Q(x,y)].
    \label{E_TrLn}
\e
In the second equality we used (\ref{G_fin}), $T$ is the (infinite) time
interval  and $S$ --- the surface area
of the defect. For the explicit calculations we first make the Fourier
transformation as in (\ref{fourier}). Then
\be {\cal E}
    =\mu^{4-d}\,\int\frac{d^{d-1}\vp}{2(2\pi)^{d-1}}\mbox{Tr}\ln[Q(\vp; x_3,y_3)],
    \label{E_TrLn2}
\e
where we introduced dimensional regularization and an auxiliary
mass parameter $\mu$ to handle the UV-divergencies.

Using the definitions of $U=U(\kappa)$ and $Q=Q(\kappa)$ we can express
the $\kappa$-derivative of the integrand of (\ref{E_TrLn2}) as
$
    \partial_\kappa \ln Q={\cal D}_{E^2} W =-{U}{\kappa}^{-1}
$.
Then for the energy density we get
\be
\label{res1}
    {\cal E}
    =-\mu^{4-d}\,\int_0^\kappa \frac{d\kappa'}{\kappa'}
        \int\frac{d^{d-1}\vp}{2(2\pi)^{d-1}}\mbox{Tr}\, U(\kappa').
\ee
We have chosen the lower limit of integration over $\kappa'$ to
satisfy the energy normalization condition
${\cal E}|_{\kappa=0}=0$. As we show below the integral is convergent at
$\kappa'=0$.

The trace of the integral operator $U(\kappa)$ is straightforward \be
\mbox{Tr}\, U\equiv \int_{-\ell/2}^{\ell/2} dx\, U(x,x)
    =2 b\ell+\frac{4 a \sinh(\ell\sqrt\rho)-\ell\kappa}{2\sqrt\rho}
    \label{Tr U}
\ee where we already used that $a=c$. Using $a$ and $b$ given in
(\ref{abc}), one easily notes that $\mbox{Tr} U\sim - \ell
\kappa/(2 E)$ when $\kappa\to0$, thus supporting the above
statement.

Next, putting (\ref{abc}) into (\ref{Tr U}) we can prove directly
that \be \mbox{Tr} U
    =-\kappa\frac{\partial}{\partial\kappa}
        \ln\left[ \frac{e^{-\ell(E+\sqrt\rho)}}{4E\sqrt\rho\xi}\right].
    \label{Tr U1}
\ee Thus, from (\ref{res1}) and  (\ref{Tr U1}) we obtain the
following expression for the Casimir energy
\be
{\cal E}=\mu^{4-d}\int \frac{d^{d-1}\vp}{2 (2\pi)^{d-1}}
    \ln\[
        \frac{e^{- \ell E}}{4 E \sqrt\rho}
            \(e^{ \ell\sqrt\rho }(E+\sqrt\rho)^2-e^{- \ell\sqrt\rho }(E-\sqrt\rho)^2\)
    \]
    \label{E}
\ee
For the first time the explicit results for the Casimir energy
of single slab were obtained in \cite{Bordag'95}, and recently
rederived in \cite{Vass Kon 08}. Upon using different
regularization scheme and notation they coincide with (\ref{E}).
The question of comparison of  results obtained in different
regularization schemes  is addressed in \ref{Reg depen}.

To extract the UV divergencies at $d=4$, we break up the energy in
two parts \be
    {\cal E}= {\cal E}_{\it fin} + {\cal E}_{\it div},
\ee where \be {\cal E}_{\it fin}= \frac1{4 \pi^2}\int_0^\infty
\Xi(p)
    p^2 dp , \label{E_fin}
\ee
$$
\Xi(p)\equiv
    \ln\[
        \frac{e^{-2 \ell E}}{4 E \sqrt\rho}
            \(e^{2\ell\sqrt\rho }(E+\sqrt\rho)^2-e^{-2\ell\sqrt\rho }(E-\sqrt\rho)^2\)
    \]
    -\frac{\lambda}{2E}\(1-\frac{\lambda}{4 \ell E^2}\),
$$
\be {\cal E}_{\it div}=\frac{\lambda\mu^{4-d}}{2 (2 \pi)^{d-1}}
    \int \frac{d^{d-1} p}{2E}\(1-\frac{\lambda}{4 \ell E^2}\).
    \label{E_div}
\ee
It easy to check that the first item,  ${\cal E}_{\it fin}$,
is finite if  we remove regularization,  while ${\cal E}_{\it
div}$ is divergent. Its asymptotic in the limit of removal of
regularization is given by
\begin{eqnarray}
  {\cal E}_{div} &=& \frac{\lambda(2\ell m^2+\lambda)}{32 \pi^2\ell (d-4)}+
    \label{E div assy}\\
  &&\frac\lambda{64 \pi^2\ell}
        \(
        \lambda
        +(2\ell m^2+\lambda)\(\gamma_E-1-\ln(4\pi)+2\ln\frac{m}\mu\)
        \)
    +\Or(d-4)\nonumber
\end{eqnarray}
Here $\gamma_E=0,577215$ is the Euler constant.
Explicit calculation of ${\cal E}_{\it div}$
will be used in the considerations of Section \ref{no div}.

\subsection{Renormalization procedure}
Simple analysis of the dependence of ${\cal E}_{\it div}$
(\ref{E_div}) on the parameters $\lambda$, $\ell$ shows that for
the renormalization of the model  at the one-loop level considered
here we must add to the action \textit{at least} the following
field-independent counter-term $\delta S$
\be
    \delta S = f \lambda   + g \lambda^2 \ell^{-1},
    \label{c-terms}
\ee
with bare parameters $f$ and $g$ (of the mass dimensions two and zero
correspondingly)
\footnote{ We remind the reader that we consider a free scalar
field, thus the counter-term $\delta S$ is the only one
required.}.
It allows us to choose these parameters in such a way that the
renormalized Casimir energy ${\cal E}_{r}$ defined by the full
action $S+\delta S$ and considered as the function of the renormalized
parameters appears to be finite both in the regularized theory, and
also after the removing of the regularization.

Thus, within such `minimal addition' renormalization scheme we
obtain for the renormalized Casimir energy the following result
\be
    {\cal E}_r= {\cal E}_{\it fin} + \lambda f_r + g_r {\lambda^2 }{\ell}^{-1}
    \label{E_ren}
\e
where finite  parameters $f_r$, $g_r$ must be determined with
appropriate experiments, or fixed with the normalization conditions.
The number of the required conditions is determined by the
(in)dependence of the coupling constant $\lambda$ on the slab
thickness $\ell$. However, in the most interesting cases
considered below only one normalization condition is necessary.

Indeed, let us consider the Casimir pressure. It is the observable quantity defined by
\be
    p=-\frac{\partial {\cal E}_r}{\partial \ell}.
    \label{p_cas}
\ee
For constant $\lambda$ it leads to
\be
    p_{\it matt}=-\frac{\partial {\cal E}_{\it fin}}{\partial \ell}+
            \frac{g_r\lambda^2}{\ell^{2}}.
        \label{p_cas_matt}
\ee
where only one renormalization parameter is present.

\begin{figure}
\centerline{ \psfig{figure=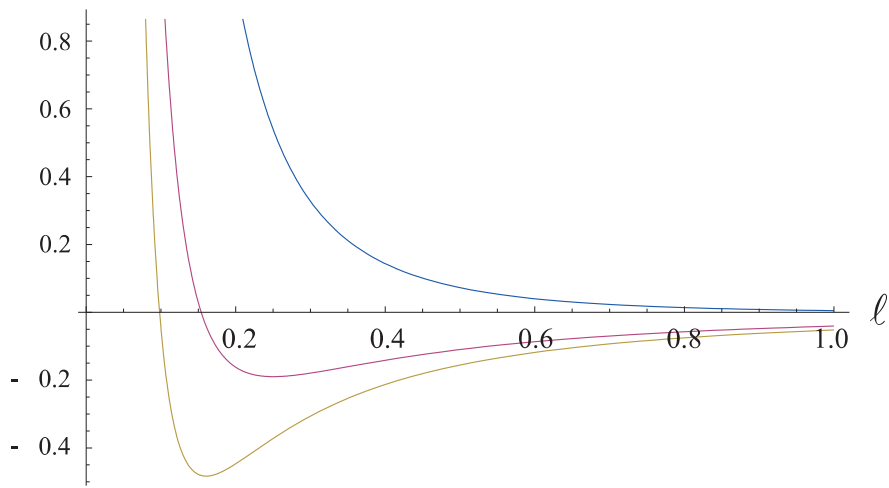,height=2.5in}} \caption{
    Renormalized Casimir pressure  $p_{\it matt}$ (\ref{p_cas_matt})
    as a function of $\ell$ for $\lambda=3$,
    $4 \pi^2 g_r =-0.1$, $-0.3$, $-0.35$ (top to bottom)
    in units of $m=1$.
    }
\label{e_cas pict}
\end{figure}

The original definition (\ref{theta}) of the distribution function
$\theta(\ell, x_3)$ corresponding to this case can be interpreted
as preserving the amount of matter
within the slab: $\int dx_3 \theta(\ell, x_3)=1$. However,
$g_r$ remains up to the  experimental fixing since there is
no other natural normalization condition that could be used in this case, unlike the case discussed in
the following section.
The plot of the
$p_{\it matt}$ as a function of $\ell$ for different values of
$g_r$ is presented at Fig. \ref{e_cas pict}. Note the
non-monotonous dependence of the pressure on the width of the slab
for comparatively large negative $g_r$.

\subsection{Dirichlet limit}\label{Dir lim Sect}
Alternatively, one can consider the density of the matter to be
fixed and calculate the pressure under this condition. Then the
distribution function has a different normalization condition
$\int dx_3 \theta(\ell, x_3)=\ell$, which is equivalent to the
change of variables  $\lambda\to\ell \kappa$ in the formulae
(\ref{E_fin})--(\ref{E_ren}). In this case the two counter-terms
in (\ref{E_ren}) can be effectively combined into a single one
$\tilde g_r$ of mass dimension one
$
    {\cal E}_r= {\cal E}_{\it fin} + \kappa \ell \tilde g_r,
$
which gives for the pressure
\be
    p_r= -\frac{\partial{\cal E}_{\it fin}}{\partial \ell}- \kappa \tilde g_r.
        \label{p_ren1}
\ee
For fixing $\tilde g_r$ we can use what we a call a Dirichlet
limit, that also provides a natural way to establish a
correspondence between our results and the previous calculations.

One can note that putting
$    \kappa=-m^2
$
and then taking the $m\to\infty$ limit effectively converts
the system under consideration into a massless scalar field
confined within a finite interval in $x_3$:
$x_3\in(-\ell/2,\ell/2)$ subject to Dirichlet boundary conditions
at the endpoints.

To investigate the behavior of the finite part of the Casimir
energy ${\cal E}_{\it fin}$ in this limit, one has to construct  carefully  the $m\to\infty$
asymptotics of (\ref{E_fin}). It can be done rigourously with help of
the Taylor expansion of the integrand of (\ref{E_fin}) in powers of
$e^{-2mp\ell}$ and consequent resummation of the same order contributions.
This procedure yields
\be
   {\cal E}_{\it fin}=-\frac{m^4\ell}{128 \pi^2}
        +\(\frac\pi{6}-\frac{4}{9}\)\frac{m^3}{4 \pi^2}
        -\frac{\pi^2}{1440 \ell^3}+\Or(m^{-1})
   \label{E fin Dir}
\ee
Now we can require that the renormalized Casimir pressure (\ref{p_ren1}) in
this limit coincide with one calculated for the case of a
massless scalar field subject to the Dirichlet boundary conditions
\cite{Milton-BOOK'01}
\be
    p_{\it Dir}=-\frac{\pi^2}{480 \ell^4}.
\ee
This condition fixes the renormalization parameter $\tilde g_r$ of
(\ref{p_ren1})
\be
    \tilde g_r=-\frac{m^2}{128 \pi^2}.
    \label{t g_r}
\ee

The Dirichlet limit procedure presented here looks somewhat
similar to the `large-mass prescription' widely used in
the calculations of the Casimir energy \cite{Bordag-Mohideen-Mostepanenko-OBZOR'01}.
However, the two schemes
have different physical motivation. While in the large mass
prescription  it is argued on some general grounds that the Casimir
energy must vanish in the large mass limit, in our case we
collate a particular limit of our results with a well
known (unambiguous) physical situation.

\subsection{Vacuum Instability}

It is well known that quantum systems might become unstable for
sufficiently deep negative potentials. In our case the threshold
is given by $\lambda=-\ell m^2$ and for deeper potentials the
Casimir energy (\ref{E_fin}) acquires an imaginary part \cite{Bordag'95}. However
the generation of the imaginary part is not only regulated by the
strength of the potential but also by its width. Moreover, the
stability of the systems depends on the separation $\ell$ in a
non-monotonous way.

\begin{figure}
\centerline{ \psfig{figure=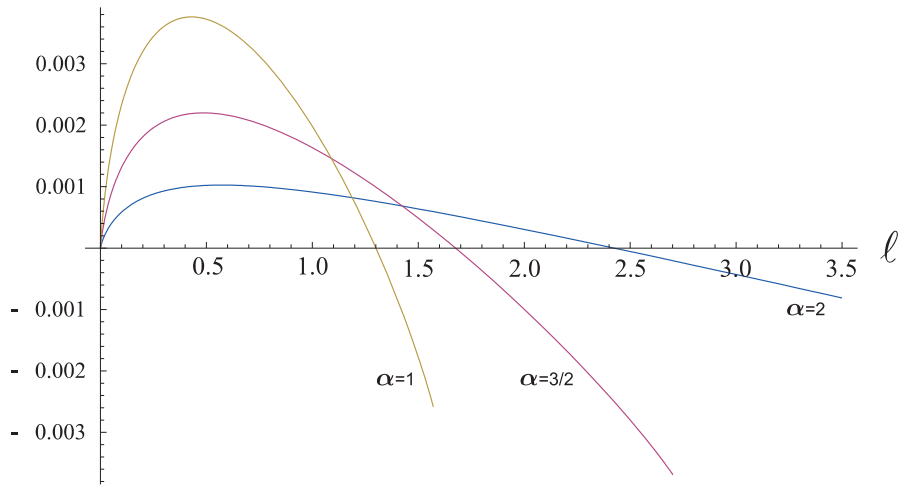,height=2.5in}} \caption{
    Casimir energy ${\cal E}_{\it fin}$ (\ref{E_fin})
    as a function of $\ell$ for $m=1$, $\kappa=-\alpha m^2 $
    and different $\alpha$-s.
    }
\label{alphas pict}
\end{figure}

From the technical point of view the imaginary part of the Casimir
energy (\ref{E_fin}) is due only to the negative values of the
argument of the logarithm in $\Xi(p)$. It is clear that $\Xi(p)$
is still real for $\kappa/m^2\equiv -\alpha >-1$. However, for
bigger values of $\alpha$: $\alpha>1$, the argument of the
logarithm is not positively defined anymore in the region of small
$p$: $p<m\sqrt{\alpha-1}$. As a function of the separation it changes
sign at every $\ell_0$ satisfying
\be
    \ell_0=\frac1{\sqrt{|\rho|}}
    \(\pi-\arctan\frac{2E\sqrt{|\rho|}}{E^2 - |\rho|}\).
    \label{ell0 sol}
\ee
The maximum separation at which the Casimir energy is still
real is given by the minimum of $\ell_0$ as a function of $p$. It
is acquired at $p=0$ and we can conclude that the Casimir energy
is real for all separation $\ell$ less or equal to
\be
    \ell_0
        = \frac{1}{m\sqrt{\alpha-1}}\(\pi-\arctan\frac{2\sqrt{\alpha-1}}{2-\alpha}\).
\ee
while for larger $\ell$ it is complex. The dependence of the
Casimir energy on the separation $\ell$ for $\ell<\ell_0$ is
presented on Fig.~\ref{alphas pict}.

\subsection{Absence of divergences}\label{no div}
A particulary interesting case of an unstable system is given by
$\lambda=-2\ell m^2$ (or $\alpha=2$ in notation of the previous
subsection). It is characterized by the absence of divergencies as
it immediately follows from (\ref{E div assy}). It is possible to
check that this value of $\lambda$ corresponds to the case of
vanishing second heat kernel coefficient $a_2$ --- the situation
known to possess no divergencies \cite{Blau Visser Wipf 88}.
Note that in this case the
dependence on the auxiliary parameter $\mu$ also disappears.

Using the condition elaborated in the previous section we conclude
that ${\cal E}_{Cas}$ acquires an imaginary part for $\ell
m>\pi/2$. It is pedagogical to rewrite this condition in CI units
$$
    \ell > \frac\pi2 \frac\hbar{mc}
$$
then at the r.h.s. we recognize immediately the Compton wave
length (up to numerical factor) of the particle corresponding to
the quantum field. Thus, at the separations (much) larger then the
Compton wave length the Casimir energy becomes complex, and the
particles are created in the potential well. Such property has
a clear semi-classical interpretation --- at the separations less then the
Compton wave length, there is merely not enough space between the
semi-infinite slabs to host any particles.

In the opposite limit --- at short separation, --- one can derive
the following asymptotics for the Casimir energy
\be {\cal E}_{Cas}=
    \frac{m^3}{4\pi^2}
    \(
    -x\(\gamma_E-\frac1{8}+\ln x\)
        +3 x^2\ln x
    \)
    +\Or(x^2),\qquad x\equiv\ell m
\ee
It reveals that for small separations: $x\lesssim\tilde
x\equiv\tilde\ell m\approx0.473$ the potential is attractive. At
$\tilde x$ the energy has an absolute maximum, the force
$F=-m\partial_x {\cal E}_{Cas}$ vanishes (at the $\Or(x^2)$
level), and the system has an unstable equilibrium position.
With larger $x>\tilde x$ (but still $x<\pi/2$) the Casimir
interaction give rise to a repulsive force on the semi-infinite
slabs. This behavior has particular similarities to one described
in \cite{Blau Visser Wipf 88} for negative second heat kernel
coefficient $a_2<0$. However, we remind the reader, that in our
case $a_2$=0.
This analytical study is supported by numerical evaluation of the
Casimir energy presented at the Figure \ref{alphas pict}.

Summarizing, we can say that at small distances $\ell<\tilde\ell$, the
gap between the slabs tends to shrink thus restoring the
homogeneity of the space filled uniformly with matter. If
additional energy is brought to the system be moving the slabs to
the distances $\ell>\tilde\ell$ the system becomes unstable emitting
particles and the gap growing. This can also be thought as
restoring the most homogenous state of the space-time filled uniformly
with the particles.

\section{Interaction of two finite width mirrors}\label{2L}
\subsection{The statement of the problem}
The method developed above does not only give in a closed form
the main characteristics of a QFT model with a single material slab,
but can also be generalized for more complicated geometries.

In particular, let us consider two plane slabs of thickness
$\ell_{1,2}$ interacting over the distance $r$. In this case the
action can be written as
\begin{eqnarray}
  S &=& S_0+S_{\it def}  \label{action 2L}\\
    && S_0=\frac12\int d^4x\, \phi(x) (-\partial^2_x+m^2) \phi(x)
        \nonumber\\
    &&
    S_{\it def}=
        \int d^3x \(\kappa_2\int_{-a_2-\ell_2}^{-a_2} dx_3 \phi^2(x)
            +\kappa_1\int_{a_1}^{a_1+\ell_1} dx_3\phi^2(x)\)
       \nonumber
\end{eqnarray}
we assume $\kappa_i$ independent of $x_3$ but corresponding
generalization is possible, $a_1+a_2\equiv r$.

To obtain a formal expression for the generating functional $G[J]$ (\ref{G(J)}) with
$S$ given above one can proceed in one of the following
ways. In a most direct approach, one introduces auxiliary fields
$\psi_{1,2}$ following (\ref{psi}) separately for each of the
slabs, and integrate consequently over $\phi$, $\psi_1$, and then
over $\psi_2$. Alternatively, one can consider the system with one
slab as an initial `free system', and just substitute  in final
formulae (\ref{G_fin}) and (\ref{Qx}) the free
field propagator $D$ with $\hat S$ (\ref{hat S}). With the knowledge of the
explicit formulae for $\hat S$ the later approach is much easier,
and the results of both of them naturally coincide
\be
G[J]=[{\rm Det} {\cal V}_2]^{-1/2}
    e^{\frac12J\hat S_{2L} J},
\e
\be
{\hat S}_{2L}={\hat S}
    -\kappa_2 \({\hat S} {\cal O}_2\){\cal V}_2\({\cal O}_2{\hat S}\)
 \label{S_2L}
\e
\be
    {\cal V}_2=\textbf{1}+\kappa_2 P_2,
    \label{V_2}
\e
\be
   P_2
    = {\cal O}_2 {\hat S} {\cal O}_2
   \label{P_2}
\ee
here ${\hat S}=D-\kappa_1 (D{\cal O}_1) Q_1^{-1} ({\cal O}_1
D)$ is the propagator of the free scalar field in the presence of
solitary slab number 1, $\kappa_i$, ${\cal O}_i$ and $Q_i$, as
well as other notations used below have the same meaning as in
Sect. \ref{single} but related to the slab denoted by the  subscript.

The linear (integral) operators ${\cal V}_2\equiv{\cal V}_2(x,y)$,
$P_2\equiv P_2(x,y)$ are defined within the support of the layer
$2$. Accordingly, in what follows we assume that the free
spatial arguments of the Fourier transformation of the integral
operators belong to $(-a_2-\ell_2,-a_2)$.

\subsection{Casimir energy}\label{2L en sect}
The energy density per unit area of the layers is given by \be
{\cal E}
    =\mu^{4-d}\,\int\frac{d^{d-1}\vp}{2(2\pi)^{d-1}}
        \mbox{Tr}\ln[{\cal V}_2(\vp; x_3,y_3)].
    \label{E_TrLn-dbl}
\e
For the explicit calculation of the energy we follow the technique
developed in Sect. \ref{E single Sect} and introduce operator $J_2$
\be
    J_2={\cal V}_2^{-1}-\textbf{1}.
\ee
The defining property of $J_2$
can be expressed via (\ref{V_2}) with the following operator equation
\be
    J_2 + \kappa_2 P_2(1+J_2)=0.
    \label{J_2 def}
\ee
According to (\ref{P_2}) $P_2$ does not depend on $\kappa_2$, which
allows us to express the $\kappa_2$-derivative of the integrand of
(\ref{E_TrLn-dbl}) in the following form
$$
    \frac{\partial}{\partial \kappa_2} \ln{\cal V}_2
        =\frac{P_2}{{\cal V}_2}=P_2 (1+J_2)=-\frac{J_2}{\kappa_2}
$$
and for the Casimir energy density we obtain in complete analogy with
the case of one slab
\be {\cal E}
    =-\mu^{4-d}\,\int_0^{\kappa_2} \frac{d\kappa_2}{\kappa_2}
        \int\frac{d^{d-1}\vp}{2(2\pi)^{d-1}}\mbox{Tr} J_2.
\ee

To solve (\ref{J_2 def}) and find the operator $J_2$ explicitly
we first use the expression of the
propagator ${\hat S}$ (\ref{hat S})
to calculate $P_2$ (\ref{P_2}). Assuming
that slabs do not intersect $r\geq \epsilon>0$ we employ (\ref{S --})
and derive
\be
    P_2(x,y)={\cal D}_{E^2}(x,y)+ c_1 e^{E(x+y)}
\ee
$$
    c_1=\frac{\kappa_1\xi_1}{2 E}e^{-2a_1 E}(1-e^{2\ell_1\sqrt\rho_1})
$$

Taking this expression into account we deduce that $J_2$ possesses
properties analogous of those of $U$ (\ref{s1})
and satisfies the following equation
$$
    K_\rho J_2= -\kappa_2.
$$
We search for $J_2$ as a sum of partial solution of
this equation and general solution of its homogeneous version
$$
J_2(x,y)=-\kappa_2 {\cal D}_{\rho}(x,y)
    +A e^{(x+y) \sqrt{\rho_2}}+B\left(e^{(x-y) \sqrt{\rho_2}}+e^{(y-x) \sqrt{\rho_2}}\right)
        +C e^{-(x+y) \sqrt{\rho_2}}.
$$
To deduce the coefficients $A$, $B$, $C$ we require that (\ref{J_2
def}) is identically satisfied for all $x, y$ belonging to
$(-a_2-L_2,-a_2)$, then
\begin{eqnarray}
  A &=& -\frac{\kappa_2 \zeta_2}{2\sqrt{\rho_2}}
    e^{2(a_2+\ell_2)\sqrt{\rho_2}}
        \(\kappa_2 e^{2 a_2 E}+2 c_1 E(E+\sqrt{\rho_2})^2\),
    \label{ABC}\\
  B &=& -\frac{\kappa_2 \zeta_2}{2\sqrt{\rho_2}}
    \( e^{2 a_2 E}(E-\sqrt{\rho_2})^2+2 c_1\kappa_2 E\),
     \nonumber\\
  C &=& -\frac{\kappa_2 \zeta_2}{2\sqrt{\rho_2}} e^{-2a_2 \sqrt{\rho_2}}
    \(\kappa_2 e^{2 a_2 E}+2 c_1 E(E-\sqrt{\rho_2})^2\),
    \nonumber
\end{eqnarray}
here $\zeta_2$ is defined through
\be
    \zeta_2^{-1}=\xi_2^{-1} e^{2 a_2 E}+2 c_1\kappa_2 E (e^{2 \ell_2\sqrt{\rho_2}}-1).
\ee

For the trace of $J_2$ we obviously have
\be
\textrm{Tr}J_2\equiv\int_{-a_2-\ell_2}^{-a_2} J_2(x,x)dx=
    \ee
    $$
    =2\ell_2 B
    -\frac{1}{2\sqrt{\rho_2}}\(
    \ell_2\kappa_2-
        A e^{-2a_2\sqrt{\rho_2}}(1-e^{-2\ell_2\sqrt{\rho_2}})
        +C e^{2a_2\sqrt{\rho_2}}(1-e^{2\ell_2\sqrt{\rho_2}})
    \)
$$
Using explicits for $A$, $B$, $C$ (\ref{ABC})
for the energy we derive
\be
{\cal E}
    =-\mu^{4-d}\,\int_0^{\kappa_2} \frac{d\kappa_2}{\kappa_2}
        \int\frac{d^{d-1}\vp}{2(2\pi)^{d-1}} \varepsilon_2,
\ee
$$
\varepsilon_2 =
        -\frac{\kappa_2\ell_2}{2\sqrt{\rho_2}}
            \(1+2\zeta_2[e^{2 a_2E}(E-\sqrt{\rho_2})^2+2c_1\kappa_2 E]\)
            $$
            $$
        \hskip2cm+\frac{\kappa_2\zeta_2}{2 \rho_2}(1-e^{2\ell_2\sqrt{\rho_2}})\(\kappa_2e^{2 a_2E}+2 c_1E(2E^2+\kappa_2)\).
$$
This somewhat cumbersome expression can be substantially
simplified since the integration over $\kappa_2$ can be made
explicitly. This also makes explicit the $1\leftrightarrow2$ symmetry of
the layers. One obtains
$$
{\cal E}_{2L}=\mu^{4-d}\int\frac{d^{d-1}\vp}{2(2\pi)^{d-1}}\times
    $$
    $$
    \times
        \log
        \[
            \frac{
            e^{-\ell_1(E+\sqrt{\rho_1})-\ell_2(E+\sqrt{\rho_2})}
            \(1-\kappa_1\kappa_2\xi_1\xi_2
                e^{-2 E r}
                (1-e^{2 \ell_1\sqrt{\rho_1}})
                (1-e^{2 \ell_2\sqrt{\rho_2}})\)
            }
            {16 E^2 \xi_1\xi_2\sqrt{\rho_1}\sqrt{\rho_2}}
        \]
$$
which finally is rewritten as
\be {\cal E}_{2L}
    ={\cal E}_{1}+{\cal E}_{2}
    +\int\frac{d^3\vp}{2(2\pi)^3}
        \log
        \[1-e^{-2 E r}\prod_{i=1,2}\kappa_i\xi_i
                (1-e^{2 \ell_i\sqrt{\rho_i}})
        \].
    \label{E 2L}
\ee
Here ${\cal E}_{1,2}$ give the self-energy (\ref{E}) of the solitary layers
$1,2$ correspondingly. The third term in (\ref{E 2L}) represents
the interaction of two layers and vanishes in the limit
$r\to\infty$. Taking into account the $p\to\infty$ behavior of
$\xi_i$ (\ref{abc}) one notes that the interaction term is UV finite, and
the removal of regularization made in (\ref{E 2L}) is indeed
justified. This is in perfect accordance with the general
considerations \cite{Emig 07} of the finiteness of the Casimir interaction
between disjoint bodies. Thus, the interaction between the slabs is
given by unambiguously finite force
\begin{eqnarray}
  {\cal F}_{2L} &\equiv&
    -\frac{\partial {\cal E}_{2L}}{\partial r} \nonumber\\
    &=&-\int\frac{d^3\vp}{(2\pi)^3}\,
        \frac{E \prod_{i=1,2}\kappa_i\xi_i
                (1-e^{2 \ell_i\sqrt{\rho_i}})}
        {e^{2 E r}-\prod_{i=1,2}\kappa_i\xi_i
                (1-e^{2 \ell_i\sqrt{\rho_i}})
        }.
        \label{F 2L}
\end{eqnarray}

\subsection{Interaction of a single slab with a delta-spike}\label{Limits sect}

Basing on the general formulae for the Casimir interaction of two
layers several limiting cases could be considered.

First of all, the limit $\ell_2\to0$ with
$\kappa_2=\lambda_2/\ell_2$, $\lambda_2$ fixed, brings the
interaction of a delta-function spike with a finite-width slab
\be
{\cal E}_{\delta L}
    ={\cal E}_{1}
    +\int\frac{d^3\vp}{2(2\pi)^3}
        \log
        \(1-\frac{e^{-2 E r}\kappa_1\xi_1\lambda_2}{2 E +\lambda_2}
                (1-e^{2 \ell_1\sqrt{\rho_1}})
        \).
    \label{E delta L}
\ee
It is worth mentioning that in this system there are two
(potentially) observable quantities: the self-pressure of the slab in
presence of a delta-spike, and the interaction between two of them.
According to this, we retained in (\ref{E delta L}) the
self-energy of the first layer ${\cal E}_{1}$, while discarded the
self energy of delta-spike which is geometry independent
and by no means is observable. Apart
from this, we note that no trace of any `sharp limit'
complications \cite{Graham Jaffe 04} arise in this situation. The interaction of a
finite width slab and an (infinitely) thin one is not influenced
in any sense by the divergencies associated with the limit
$\ell_2\to0$.

Now we can further consider in (\ref{E delta L}) the limit $\ell_1\to0$ with
$\kappa_1=\lambda_1/\ell_1$, $\lambda_1$ fixed,
to prove that (\ref{E 2L}) does lead to the standard result for
the Casimir interaction of two delta-spikes separated by the
distance $r$ \cite{Bordag-Mohideen-Mostepanenko-OBZOR'01}
\be
    {\cal E}_{2\delta}=
         \int\frac{d^3\vp}{2(2\pi)^3}
         \log\(
         1-\frac{e^{-2Er}\lambda_1\lambda_2}{(2 E+\lambda_1)(2 E+\lambda_2)}
         \),
\ee
where again the distance independent contribution was omitted.

\begin{figure}
\centerline{ \psfig{figure=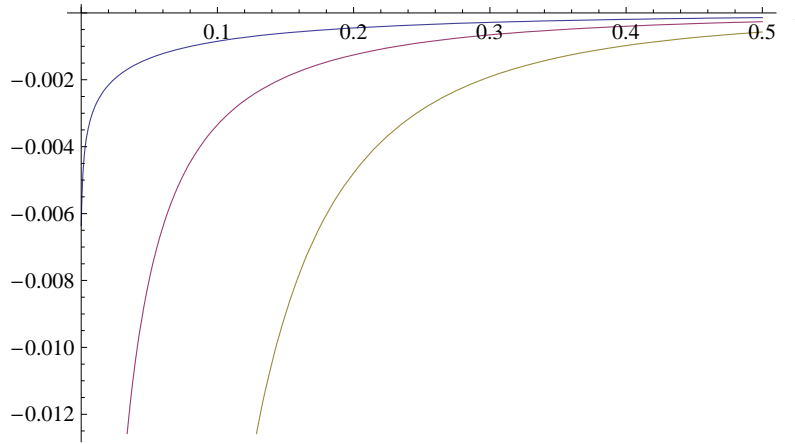,height=2.5in}}
\caption{
    Casimir force between two layers, layer and delta-spike,
    two delta-spikes (from top to down) as function of separation $r$,
    for $m=0.1$, $\lambda_1=\lambda_2=2$, $\ell_1=\ell_2=1/2$ when appropriate.}
\label{f_two obj}
\end{figure}

The force acting between two distinct objects as function of
separation is plotted on the Fig. \ref{f_two obj} for the
following cases: two finite-width layers, finite-width layer and
delta-spike, two delta-spikes. The most interesting peculiarity of
two slabs interaction is that it remains finite in the limit
$r\to0$, which can also be seen at the analytical level.

\subsection{Connection to the Lifshitz formula}\label{Lifsh sect}
It is also possible to consider the opposite limit of the slabs of infinite width separated by
the finite distance $r$ (i.e. the interaction between two semi-infinite material slabs).

In this case we put
$a_1=a_2=r/2$, $\ell_1=\ell_2=\ell$, $\lambda_i=\ell \kappa_i$,
and consider $\ell$ tending to infinity: $\ell\to\infty$ while
keeping $\kappa_i$ finite. Then basing on ${\cal E}_{2L}$ (\ref{E 2L})  we derive
\be
  {\cal E}_{\it Lif}
    =\int\frac{d^3\vp}{2(2\pi)^3}
        \log
        \[1-e^{-2 E r}\frac{\kappa_1 \kappa_2} {(E+\sqrt{\rho_1})^2 (E+\sqrt{\rho_2})^2}\]
    \label{E l1}
\ee
where the $r$-independent (divergent) terms are omitted.

Taking into account that
\be
    \rho_{1,2}=E^2+\kappa_{1,2}
\ee
we can rewrite (\ref{E l1}) further as
\be
  {\cal E}_{\it Lif}
    =\int\frac{d^3\vp}{2(2\pi)^3}
        \log
        \[1-e^{-2 E r}\frac{(E-\sqrt{\rho_1}) (E-\sqrt{\rho_2})} {(E+\sqrt{\rho_1}) (E+\sqrt{\rho_2})}\].
    \label{E l2}
\ee
This gives for the force per unit area of the semi-infinite slabs
\begin{eqnarray}
  {\cal F}_{\it Lif} &\equiv&
    -\frac{\partial {\cal E}_{\it Lif}}{\partial r} \nonumber\\
    &=&-\int\frac{d^3\vp}{(2\pi)^3}\,
        \frac{E}
        {e^{2 E r} \frac{(E+\sqrt{\rho_1}) (E+\sqrt{\rho_2})} {(E-\sqrt{\rho_1}) (E-\sqrt{\rho_2})}-1}.
        \label{F lif}
\end{eqnarray}
This formula is somewhat similar to (3.8) in
\cite{Milton-OBZOR'04} which represents the contribution of the TE
modes of the electromagnetic field to the force acting between two
semi-infinite dielectric slabs. One notes that the correspondence
becomes exact if in our model we introduce particular dispersion
into the interaction of quantum fields with the material defect,
namely
\be
    \kappa_{1,2}^{TE}\equiv \kappa_{1,2}^{TE}(p)=(\epsilon_{1,2}-1) p_0^2.
        \label{kappa_te}
\ee
This is formally equivalent to substitution  of standard
dispersion relation $E=\sqrt{p_0^2+\vec{p}^2}$ with the new ones,
$E_i=\sqrt{\epsilon_i p_0^2+\vec{p}^2}$, valid inside the corresponding
slabs. The new parameters $\epsilon$ can be considered as the
dielectric permittivity of the matter constituting them.

Since  the contribution from the TM modes can be obtained by replacement
\cite{Milton98}
\be
   E=\sqrt{\vec{p}^2+ p_0^2}
    \to \frac{\sqrt{\vec{p}^2+\epsilon p_0^2}}{\epsilon}
\ee
we find that the following dispersion
\be
    \kappa_{1,2}^{TM}\equiv \kappa_{1,2}^{TM}(p)
    =\(\frac{1}{\epsilon_{1,2}^2}-1\) \vec{p}^2+\(\frac{1}{\epsilon_{1,2}}-1\) p_0^2.
    \label{kappa_tm}
\ee
is to be introduced into the  coupling constant to obtain the TM contribution to the dielectric slabs interaction.
Introducing (\ref{kappa_te})  and  (\ref{kappa_tm}) into
(\ref{F lif}) and summing both contributions we immediately recover
the celebrated Lifshitz formula \cite{Lifshitz'56}
\be
  {\cal F}_{\it Lif} =
    -\frac{1}{4 \pi^2} \int_0^{\infty} dp_0 \int_0^{\infty} d p^2 E (d_{TE}^{-1}+d_{TM}^{-1})
\ee
\be
   d_{TE}   =  {e^{2 E r} \frac{(E+\sqrt{\rho^{TE}_1}) (E+\sqrt{\rho^{TE}_2})} {(E-\sqrt{\rho^{TE}_1}) (E-\sqrt{\rho^{TE}_2})}-1}
\ee
\be
    d_{TM}   =  {e^{2 E r} \frac{(E+\sqrt{\rho^{TM}_1}) (E+\sqrt{\rho^{TM}_2})} {(E-\sqrt{\rho^{TM}_1}) (E-\sqrt{\rho^{TM}_2})}-1},
\ee
where
\be
\rho^{TM}_{1,2}=E^2+\kappa^{TM}_{1,2}(p), \qquad
    \rho^{TE}_{1,2}=E^2+\kappa^{TE}_{1,2}(p).
\ee

The standard form of the Lifshitz formula in terms of
$\tilde s_{1,2}$ and $\tilde p$ (we use the tilde to
distinguish the Lifshitz variables) is restored with the following substitution
\be
    \tilde p=\sqrt{\frac{\vec p^2}{p_0^2}+1} , \quad
    \tilde s_1=\sqrt{\epsilon_1-1+\tilde p^2}, \quad
    \tilde s_2=\sqrt{\epsilon_2-1+\tilde p^2}
\ee

\subsection{The self-interaction limit}\label{Self sect}
The same limiting case as considered above
provides an insight into the choice of
normalization condition  discussed in Sect. \ref{Dir lim Sect} and lead
to a better understanding of the ambiguities known in the
computations of Casimir self-interactions.

One notes that $\ell_{1,2}\to\infty$ limit taken in (\ref{action 2L}) and followed by substitution $\kappa_1=\kappa_2\to-\kappa$, $m^2\to m^2+\kappa$ effectively transforms the two-slabs system into a single-slab one. Indeed, after this substitution the separation between the mirrors itself converts into a material slab in otherwise empty space, and thus one returns to the configuration of a single slab of finite width described classically by (1). The same limiting procedure can be also performed \emph{after} calculation of the
quantum effects in two-slabs system. Then we conclude that the same physical system of a single material slab can be described within two different (independent) approaches: directly calculating quantum effects for (1), or starting first with (\ref{action 2L}) and performing the limiting procedure after quantum calculations.
For the self-consistency of mathematical description of a given physical system one would normally require the exact coincidence of results obtained in different approaches.

For the sake of comparison let us rewrite (\ref{F lif}) obtained in the infinite width limit
substituting there $\kappa_1=\kappa_2\to-\kappa$, and $m^2\to m^2+\kappa$
\be
  {\cal F}_{{\it Lif}} = \int \frac{d^3\vp}{(2\pi)^3}\,
    \frac{\kappa^2\sqrt{E^2+\kappa} }
    {\kappa^2
        -e^{2r\sqrt{E^2+\kappa} }\(E+\sqrt{E^2+\kappa}\)^4}.
    \label{F_w}
\ee
%
On the other hand,  the self-pressure calculated as $r$--derivative
of (\ref{E}) can be written as
\begin{eqnarray}
  \hskip-2cm{\cal F} &\equiv&
        -\frac{\partial \cal E}{\partial r} -\kappa \tilde g\nonumber\\
  \hskip-2cm&=&-\kappa \tilde g- \mu^{4-d}\int \frac{d^{d-1}\vp}{2 (2\pi)^{d-1}}
     \frac{\kappa\(\kappa + e^{2r\sqrt{E^2+\kappa}}\(E +\sqrt{E^2+\kappa}\)^2\)}
        {\kappa \(E -\sqrt{E^2+\kappa}\)+e^{2r\sqrt{E^2+\kappa}}\(E+\sqrt{E^2+\kappa}\)^3}.
    \label{F}
\end{eqnarray}
where we introduced explicitly the counter-term needed for renormalization of ${\cal F}$.

By direct comparison one immediately notes that  (\ref{F_w}) and (\ref{F})
differ by a
\begin{eqnarray}
  \Delta &=&-\kappa \tilde g -\frac{\mu^{4-d}}{2(2\pi)^{d-1}}\int d^{d-1}\vp
            \(E-\sqrt{E^2+\kappa}\) \nonumber\\
    &=&-\kappa \tilde g+ \mu^{4-d}\frac{\((m^2+\kappa)^{d/2}-m^d\) \Gamma(-d/2)}
            {2^{d+1}\pi^{d}}.
\end{eqnarray}
It is easy to check that $\Delta$ vanishes in the limit $d\to4$ if we fix the counter-term $\tilde g$ following the prescription
advocated in Sect. \ref{Dir lim Sect}, and the two results obtained in two different ways coincide.

Thus, we can say that two approaches describing a single-slab system
differ by the presence of divergencies. In the direct approach based on action (1)
the self--pressure (\ref{F}) appears to be formally divergent,
thus leading to an ambiguity represented by the counter-term $\tilde g$.
At the same time, the Casimir force between the two slabs
is finite as it generally should be for distinct bodies, \cite{Emig 07}, and
this property is preserved in the infinite width limit described above.
This limit transforms the systems under consideration into a single-slab one, and thus (\ref{F_w}) must also be considered as a self--pressure.  By construction it is finite and does not possess any ambiguities. For the self-consistency of our description we require exact coincidence of (\ref{F_w}) and (\ref{F}) thus fixing the value of (otherwise arbitrary) $\tilde g$. Remarkably, the same value for $\tilde g$ is obtained if we impose physical normalization condition based on the Dirichlet limit as discussed in Sect. \ref{Dir lim Sect}.

The discussed limiting procedure establishes a new divergence free approach to calculate the self--pressure of a
single finite width slab.
However, we would like to stress that this limit should not be considered as a physical process but a formal mathematical trick. Still it allows one to obtain finite unambiguous self--pressure of a single material slab.

On the other hand, one might argue that the approach based on the action (1) is actually describing a reacher physics as after renormalization the self--pressure contains an extra parameter $\tilde g$. Thus it could probably give answers to a wider range of physical questions. However, we believe that the requirement of the self-consistency should be used instead to fix the value of this parameter especially taking into account that it leads to same value of $\tilde g$ as derived in  Sect. \ref{Dir lim Sect}. Moreover, the less parameters one has in a model the more powerful are its predictions since in this case one has less fitting freedom to coincide with experiment.

The final answer discriminating between two approaches is to be given by an appropriate experiment. Such an experiment could be based on measuring the change in energy as a material slab expands or shrinks. The feasibility of such measurements can possibly be considered in condensed matter physics where the self--pressure of a (soft) material slab can be significant for the properties of the states of the systems.

\section{Conclusion}

In this paper we constructed QFT model of the scalar field interacting with the
bulk defects concentrated within slabs (mirrors) of finite width.
Within a rigorous mathematical approach we constructed basic observable quantities
for such systems and also thoroughly investigated physically interesting limiting cases.

For the system of a single mirror we calculated explicitly
the propagator, the S-matrix elements and the vacuum determinant (Casimir energy).
The latter one coincides with results obtained
previously in \cite{Bordag'95}, \cite{Vass Kon 08} within a
different approach. The explicit formulae for the full
propagator and the S-matrix elements are given for the first time.
 The Casimir energy is UV
divergent and for its regularization we applied dimensional
regularization. It allowed us to extract the finite part and to
construct the counter-terms. The renormalization procedure
requires generally two normalization conditions to fix the values
of the counter-terms with the appropriate experiments. It is shown
that the Casimir pressure in the system can be calculated in two
different ways: for fixed density of matter and for fixed amount
of matter of the slab. In the latter case we  were also able to establish a physically
sensible normalization condition by comparing the system to the well known case
of Dirichlet boundary conditions imposed on the quantum fields.  We also
considered the non-stable potentials,
when the creation of particles out of vacuum can be expected, and
investigated the properties of the divergence free type of defects.

The method used for the single mirror system proved to be applicable
to more complicated systems. In particular we considered the configuration
of two mirrors of different widths separated by the interval $r$. Using the technique
developed for the single mirror case we were able to calculate explicitly the Casimir energy
of the quantum field interacting with such defect. The system possesses two different types of
observables --- the interaction of two slabs and self-interaction of one slab influenced by the
presence  of the other. The self-interaction can also be computed in two different ways mentioned above.
Based on the general formula we considered several limiting cases. We  presented corresponding explicit
expressions and numerical evaluations of the interaction of a single slab with a delta spike
(which further reproduces correctly the limit of two delta-spikes) and interaction
between two semi-infinite slabs. Both results are presented for the first time.
Investigating the final formulae we were able to establish a direct
correspondence of obtained results with the well known Lifshitz formula
which supports the validity of our approach.

Moreover, investigating the Casimir interaction of two mirrors
we developed a formal limiting procedure which recovers in an unambiguous
and divergence free manner the Casimir \textit{self}--energy of a single slab.
As well known, the separation dependent part of the interaction between two disjoint
bodies does not possess any ultraviolet divergences. This property was found to
be transferable to the single mirror case by the developed mathematical trick thus leading
to well--defined finite self--energy.

Despite it is a unique case by now, we
believe that such limits should be searched for in other
demanded cases of Casimir self-interaction, and especially for the
case of dielectric ball.  We expect that further generalization of the  method
proposed in this paper to the case of QED and its application
to the spherical geometry is also possible.

\section*{Acknowledgement}
Authors are grateful to Prof. D. Vassilevich and Dr. V. Marachevskiy for numerous discussions
and general interest in the research.

The work was financially supported within the RFBR grant
$07$--$01$--$00692$ and RNP grant 2.1.1/1575 (I.V.F.\ and Yu.M.P.)
I.V. Fialkovsky also gladly acknowledges support of FAPESP.

\appendix
\section{Regularization dependence}\label{Reg depen}
To compare different calculations in (renormalizable) quantum
field theory one should normally operate with renormalized
quantities subject to the same normalization condition. However,
in a number of Casimir type problems (e.g. massive scalar field
 interacting with shells, etc.) the understanding of physically viable
normalization conditions is still lacking. Yet a sound comparison
is possible if the same regularization scheme is applied.

When there are different regularizations used in different
calculations, the renormalizability of considered models still
guaranties the possibility of comparison. It is due to the fact
that in renormalizable theories regularization as such is not
attributed with any independent physical meaning and different
regularization schemes must lead to the same final answer
\footnote{
We put aside an important problem of anomalies which is concerned
with breaking different symmetries by using different
regularizations. We do not discuss this issue here, as it is not
relevant for scalar fields under consideration.}.
Then an interchange of regularization schemes is possible and well
justified. In a regularized quantity one should introduce
additional regularization (which is always possible) and then
remove the original one. It is a technical problem to the
investigator to realize this programme explicitly.

To illustrate this idea, we consider our results for the Casimir
self-energy of a single slab. We can compare (\ref{E}) obtained in
dimensional regularization with previous calculations of
\cite{Bordag'95} and \cite{Vass Kon 08} where $\zeta$-function
regularization was applied, and \cite{Fosco Lombardo Mazzitelli
08} where cut-off regularizations is used.

In particular one can obtain the generalization of (8)
\cite{Bordag'95} (written for $d=3$) to arbitrary (spatial)
dimension $d$  in the form
\be V_{\textit{eff}}
    =
    -\frac12\partial_s \(
    \frac{\Gamma(d/2)\Gamma(s-d/2)}{2 \Gamma(s)}\frac{\Omega_d \mu^{3-d}}{(2\pi)^{d}}
        \sum_n (\omega_n^2+m^2)^{d/2-s}
    \)_{s=0}
\label{Veff0} \ee where $\Omega_d=2\pi^{d/2}/\Gamma(d/2)$ is the
volume of the $(d-1)$-dimensional sphere in the $d$-dimensional
space. Now we can remove safely the $\zeta$-regularization putting
$s=0$, and then using (15) of \cite{Bordag'95} we arrive at \be
V_{\textit{eff}}
    =
        \frac{\Omega_d \mu^{3-d}}{2(2\pi)^{d} }
        \int_m^\infty dk (k^2-m^2)^{d/2}
            \frac{\partial}{\partial k}\ln s_{11}(ik).
        \label{Veff}
\ee
For considered case of a single plane slab $s_{11}(ik)$  is
defined in (22) \cite{Bordag'95}. Integrating by parts in
(\ref{Veff}) and changing integration variable $k=\sqrt{q^2+m^2}$,
one can check that the right hand sides of (\ref{E}) and
(\ref{Veff}) coincide up to change of notation $L=\epsilon$,
$V_0=\eta$, $d\to d-1$. In a similar way one can verify that
result obtained in \cite{Vass Kon 08} for the plane slab in $1+1$
dimensions with help of zeta-function regularization agrees with
(\ref{E}).

Analogously we can collate our results with those presented in
Section III.B of \cite{Fosco Lombardo Mazzitelli 08}. The authors
of \cite{Fosco Lombardo Mazzitelli 08} consider (67),~(68)
\cite{Fosco Lombardo Mazzitelli 08} as being regularized by cutoff
function $\tilde\lambda$  in fixed number of dimensions $d$.
Equivalently it can be viewed as dimensionally regularized for
arbitrary (non growing) $\tilde\lambda$, and thus compared with
our result (\ref{E}). One can easily check that the massless limit
of (\ref{E}) does not coincide with (67) \cite{Fosco Lombardo
Mazzitelli 08}. As we argued in \cite{FMP Comm 08} the Casimir energy
of a single slab presented in \cite{Fosco Lombardo Mazzitelli 08}
is incorrect.

This consideration shows that when comparing results of different
calculations one should be aware of subtleties concerned  with
application of different regularization schemes. On the other
hand, it should not be considered as an insuperable obstacle to a
comparison.

\end{document}